\documentclass[preprint,12pt]{elsarticle}
\usepackage{graphicx}
\usepackage{amssymb}
\usepackage{amsmath}
\newtheorem{theorem}{Theorem}

\newtheorem{defi}{Definition}
 \newtheorem{property}{Property}
 
\journal{Journal Name}

\begin{document}
	
	\begin{frontmatter}
		
		
		%% \title{Title \tnoteref{label1}}
		%% \author{Name \corref{cor1} \fnref{label2}}
		%% \address{Address \fnref{label3}}

		\title{Control Ability with Time Attributy for Linear Continous-time Systems 
			\thanks{Work supported by the National Natural Science Foundation of China (Grant No. 61273005)}}
		
		
		\author{Mingwang Zhao}
		
		\address{Information Science and Engineering School, Wuhan University of Science and Technology, Wuhan, Hubei, 430081, China \\
			Tel.: +86-27-68863897 \\
			Work supported by the National Natural Science Foundation of China (Grant No. 61273005)}
		%	 \email{zhaomingwang@wust.edu.cn} 
		
		\begin{abstract}
			 In this paper, the control ability with time attributy for the linear continuous-time (LCT) systems are defined and analyzed by the volume computing for the controllability region. Firstly, a relation theorem about the open-loop control ability, the control strategy space (\textit{i.e.}, the solution space of the input variable for control problems), and the some closed-loop performance for the LCT systems is purposed and proven. This theorem shows us the necessity  to optimize the control ability for the practical engineering problems. Secondly, recurssive  volume-computing algorithms with the low computing complexities for the finite-time controllability region are discussed. Finally, two analytical volume computations of the infinite-time controllability region for  the systems with $n$ different and repeated real eigenvalues are deduced, and then by deconstructing the volume computing equations, 3 classes of the shape factors are constructed. These analytical volume and shape factors can describe accurately the size and shape of the controllability region.
			Because the time-attribute control ability for LCT systems is directly related to the controllability region with the unit input variables, based on these analytical expressions on the volume and shape factors, the time-attribute control ability can be computed and optimized conveniently.

		\end{abstract}
		
		
	\end{frontmatter}
	
	%	 \linenumbers
	
	\section{Introduction}
	\label{S:1}	
	
	The concepts of the state controllability and observability for the dynamical systems, born in 1960's by R. Kalman, \textit{et al}, \cite{KalmHoNar1963} became the firm basises of many branches of control theory, such as optimal control, stochastic control, nonlinear control, adaptive control, robust control, $etc$.
	The so-called state controlabiity is only a two-values logic concept whether the states in state space  can be controlled or not by the input variables of the dynamical systems, and the state observability is also a two-values logic concept whether the unmeasurable states can be estimated or not by the measurable output and input variables. The controllability concepts and corresponding criterions can tell us that the dynamical systems are controllable or not, and but 
	the control ability and control effciency can not be analyzed and got. So is observability. In fact, the quantitative concept and analysis methods for that are very important for the practical control engineering, and many control engineering problems are dying to these quantitative studies. 
	
	In past 60 years, some works stated as follows are on the quantitation of control ability.
	
	1) Based the definition of the controllability Grammian matrix $G_{c}$ and the correspongding controllability Ellipsoid $E_{c}$, in papers \cite{VanCari1982}, \cite{Georges1995}, \cite{PasaZamEul2014}, and \cite{Ilkturk2015}, the determinant value $ \det \left( G_{c} \right) $ and the minimum eigenvalue $ \lambda _{\textnormal {min} } \left (G_{c} \right ) $ of the  matrix $G_{c}$, correspondingly the volume $ \textnormal {vol} \left( E_{c} \right)$ and the minimum radius $ r _{\textnormal {min}} \left (E_{c} \right ) $ of the ellipsoid $E_{c}$, can be used to quantify the control ability of the input variable to the state space, and then be chosen as the objective function for optimizing and promoting the control ability of the linear dynamical systems. Due to lack of the analytical computing of the determinant $ \det \left( G_{c} \right) $ and eigenvalue $ \lambda _{\textnormal {min}} \left (G_{c} \right ) $, correspondingly the volume $\textnormal {vol} \left( E_{c} \right)$ and the radius $ r _{\textnormal {min}} \left (E_{c} \right ) $, these optimizing problems for the control ability are solved very difficulty, and few achievements about that have been made.
	
	2) Based the definition of the controllability region $R_{c}$, the minimum distance of the boundary $ \partial R_{c}$ to the original of the state space is used to measure and optimize the control ability \cite{ViswLongLik1984}. Because of the difficulty to compute the distance, the further works have not been reported.
	
	3) Based on the well-known PBH controllability test, a degree of modal controllability was put forward to measure and optimize the control ability with the eigenvalue mobility, \textit{i.t.}, the changing ability of the system eigenvalues by the control law  \cite{LonSirSev1982} \cite{HamElad1988} \cite{
		HamJard2014}. Because the degree is only for the single modal and not for the whole system, the further works have not been carried forward. 
	
	Recently, a systemical studies on the control ability of the input variables to the state variables for the linear discrete-time (LDT) systems are carried out by M.W. Zhao \cite{zhaomw202001} \cite{zhaomw202002} \cite{zhaomw202003} \cite{zhaomw202004}, a new quantifying concept and analysis method for the control ability are put forth. And then a relation theorem among the open-loop control ability, the control strategy space (\textit{i.e.}, the solution space of the input variables for the control problems), and the some closed-loop control performance, such as, the optimal time waste, the response speed, the robustness of the the control strategy, $etc$, is purposed and proven in paper \cite{zhaomw202003}. By the theorem in paper \cite{zhaomw202003}, we can see, optimizing the control ability of the open-loop systems is with the great significations for promoting the performance indices of the closed-loop control systems. In addtion, in these papers, the analytical expressions of the volumes and the multiple shape factors of the controllability region/ellipsoid are deduced and can be used to analyze and optimize the control ability. Based on these results, the analysis and optimization methods for the control ability of the LDT systems can be founded.
	
	In this paper, the concepts and analysis methods for control ability  of the LDT systems in \cite{zhaomw202001} \cite{zhaomw202003} \cite{zhaomw202004} will be generalized to the linear continuous-time (LCT) systems, and it is expected that the analyzing and optimizing for the time-attribute control ability of LCT systems can be carried out conveniently based on the new results in this paper.
	
	%\section{Normalization of the Variables for Comparing Control ability}
	\section{Definition on the Control Ability with Time Attribute}
	
In general, the LCT Systems can be modelled as follows:
	\begin{equation}
	\dot x_{t}=Ax_{t}+Bu_{t}, \quad x_{t} \in R^{n},u_{t} \in R^{r}, \label{eq:a0501}
	\end{equation}
	\noindent where $x_{t}$ and $u_{t}$ are the state variable and input
	variable, respectively, and matrices $A \in R^{n \times n}$ and $B \in
	R^{n \times r}$ are the state matrix and input matrix, respectively \cite{Kailath1980} \cite{Chen1998}. 
	To investigate the controllability of the
	LCT systems \eqref{eq:a0501}, the controllability matrix and the controllability Grammian matrix can be defined as follows
	\begin{align}
	P_{n} & = \left[ B, AB,\dots,A^{n-1}B \right ]
	\label{eq:a0502} \\
	G_{T} & = \int _{0} ^{T1} e^{At}B \left( e^{At}B \right)^T \textnormal {s} t \label{eq:a0503}
	\end{align}	
	That the rank of the matrix $P_{n}$/$G_{T}$ is $n$, \textit{that is}, the dimension of of the state space the systems \eqref{eq:a0501}, is the well-known criterion on the state controllability.

	In papers \cite{VanCari1982}, \cite{Georges1995}, \cite{PasaZamEul2014}, and \cite{Ilkturk2015}, the determinant value $ \det \left( G_{T} \right) $ and the minimum eigenvalue $ \lambda _{\textnormal {min} } \left (G_{T} \right ) $ of the controllability Grammian matrix $G_{T}$ can be used to quantify the control ability of the input variable to the state space, and then be chosen as the objective function for optimizing and promoting the control ability of the linear dynamical systems. Due to lack of the analytical computing of the determinant $ \det \left( G_{T} \right) $ and eigenvalue $ \lambda _{\textnormal {min}} \left (G_{T} \right ) $, these optimizing problems for the control ability are solved very difficulty, and few achievements about that were made. Out of the need of the practical control engineering, quantifying and optimizing the control ability are key problems in control theory and engineering fields, and the further studyings on that are encouraged.
	
	\subsection{The Normalization of the Input Variable and The Definitions of the Controllability Region }
	
	According to paper \cite {zhaomw202003}, to study rationally the control ability of the input variables to the state variables for the different dynamical systems, it is necessary to normalize the input variable and state variables. For studying the control ability with time attribute, the normalization of the input variablles $u_t$ is as follows
		\begin{align} U_{a} =\left\{ u_t: \Vert u_t \Vert_\infty = \max_{i \in [1,r] } \vert u_{t,i} \vert \le 1 \right\} \label{eq:a0504}
	\end{align} 
	where $r$ and $u_{t,i}$ are the input variable numbers and the $i$-th input variable of the multi-input systems, respectively. In fact, the constraint condition \eqref{eq:a0504} can be used to describe the input variables for the following two cases:
	 
	1) The input variables are bounded or with some saturation elements.
	
	2) The unit input variables are considered for conveniently computing and comparing the control ability.
	
	 Based on the above normolization on the input variables, the state controllability region and control ability with time attribute can be defined and studied. In this paper, the controllability region is a broad concept and it includes narrow controllability region and reachability region. The so-called narrow controllability region and reachability region are the state ranges of the controllable state and reachable state, respectively. The word 'controllable state' means the state that can be controlled to the original of the state space by a finite-time input sequence, and 'reachable state' means the state that can be reached form the original by a finite-time input sequence. 
	 
	 Similar to the definition of the state controllability region ( a.k.a. "recover region") for the input-saturated linear systems in papers \cite{Bernstein1995}, \cite{Hulin2001}, \cite{Hu2002}, the narrow controllability region and reachability region of the LCT Systems with the input constraints $U_a$ are defined as follows
	\begin{align} 
	R_c(T) & =\left\{x : x=\int _{0}^{T} e^{-At} Bu_t \textnormal {d} t , \;\; \forall u_t \in U_a \right\} \label{eq:a05009} \\
	R_d(T) & =\left\{x : x=\int _{0}^{T} e^{At} Bu_t \textnormal {d} t , \;\; \forall u_t \in U_a \right\} \label{eq:a05010} 
		\end{align}
		In fact, the controllability regions $ R_{c}(T)$ and reachability region $R_d(T) $ are the biggest range of the controllable states and the reachable states with the unit input variables. 
		
In paper \cite {zhaomw202001}, a special zonotope generated by the matrix pair $(A,B)$ is defined as follows
	\begin{align} 
		R_a(N) & =\left\{x : x=\sum _{k=0}^{N-1} A^{k} Bz_k , \;\; \forall z_k: \Vert z_k \Vert _\infty \le 1 \right\} \label{eq:a0501001} 
		\end{align}
and the controllability region of the LDT systems can be regarded as a special zonotope defined as \eqref{eq:a0501001}. Similarly, a smooth geometry generated by the matrix pair $(A,B)$ can be defined as follows
	\begin{align} 
R_a(T) & =\left\{x : x=\int _{0}^{T} e^{At} Bz_t \textnormal {d} t , \;\; \forall z_t: \Vert z_t \Vert _\infty \le 1 \right\} \label{eq:a0501002} 
\end{align}
Similar to the zonotope $R_a(N)$, the smooth geometry $R_a(T)$ is also a convex $n$-dimensional geometry with the origin symmetry, and can be called as smooth zonotope generated by the matrix pair $(A,B)$. 

In fact, based on the definition Eq. \eqref{eq:a0501002}, the narrow controllability region $R_c(T) $ and reachability region $R_d(T) $ can be regared as two smooth zonotopes generated by the two matrix pairs $(A,B)$ and $(-A,B)$, respectively. And then, the smooth zonotope $R_a(T) $ can be regarded as a broad controllability region. In addition, the narrow controllability region and reachability region can be transformed each other, and then some results about the geometry properties and volume computing for the two regiones can be used to generalized to another.

%	\section{Control Ability under the Unit Input Constraint}
	
	\subsection {The properties of the controllability Region $R_{a}(T) $ }
	
In paper \cite{zhaomw202003} and \cite{zhaomw202004}, some properties about the controllability region $R_a(N)$ for the LDT systems, such as, the boundary, the shape and size, are put forth and proven, and then the similar resultes for the controllability region $R_a(T)$ for the LCT systems the can be stated and proven as folows 

	\begin{property} \label{pr:pr051}
		The boundary of the controllability region $R_{a}(T) $ can be descrbed as follows 
		\begin{align}
		\partial  R_{a}(T) =
		\left\{ x\left|x=\int_{0}^{T}\textnormal{sgn}\left(d^{T} e^{At}B\right)e^{At}B \textnormal {d} t, \; \forall d\in R^{n}\right.\right\} \label{eq:a05011}
		\end{align}
	\end{property}

	\begin{property} \label{pr:pr052}
		If the LCT Systems \eqref{eq:a0501} is controllable, for a given $T>0$, the controllability region $R_{a}(T) $ is a $n$-dimensional geometry, and then for any $T_1<T_2{}$, we have
		\begin{align} 
		R_{a} (T_1) \subset R_{a}(T_2) \; \textnormal {and}\; \partial R_{a} (T_1) \cap \partial R_{a}(T_2) = \phi \label{eq:a05012} 
		\end{align}
		\textit{that is}, the geometry $R_{a}$ is strictly monotonic expansion, where $ \partial R $ is the bound of the geometry $R$. 
		
		If the systems is not controllable, for a given $T>0$, the region $R_{a}(T) $ is a $n_c$-dimensional geometry, and then for any $T_1<T_2$, we have
		\begin{align} 
		R_{a} (T_1) \subseteq R_{a}(T_2) \; \textnormal {and} \; \partial R_{a} (T_1) \cap \partial R_{a}(T_2) \neq \phi \label{eq:a05013} 
		\end{align}
		\textit{that is}, the geometry $R_{a}$ is monotonic expansion, where $n_c$ is the controllability index with the value $\textnormal {rank} P_N$. 
	\end{property}
	
\begin{figure}[htbp]
		\centering
		\includegraphics[width=0.8\textwidth]{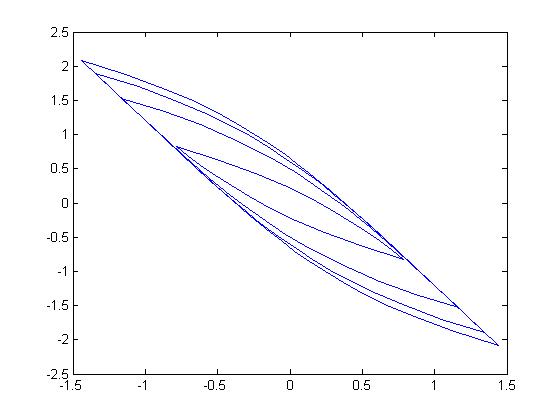}
		 %0.5指图片宽度
		\caption[c]{The 2-D   somooth zonotopes $R_a(T)$  when $T=1,1.5,3,4.5,6s$  \label {fig:aa03}}	
	
	\end{figure}

	Fig. \ref{fig:aa03}  shows the 2-dimensional   ellipsoids $R_N(S)$ generated by the matrix $(A,B)$ as follows
	$$
	A=\left[ \begin{array}{cc}
	-2.9 & -1.225 \\
	0 & -0.45
	\end{array} \right] \;\; 
	b=\left[ \begin{array}{c}
	1 \\ 1
	\end{array} \right]
	$$
	
	In the next discussion, the systems are assumed always as a controllable systems and the geometry $R_{a}$ is a $n$-dimensional smooth zonotope.
	
	\subsection {The Definition of the Control Ability with time Attributy}
	
	As we know, the bigger of the controllability region $R_{a}$ is, the more the controllable states in the state space are, and but whether the control ability of the dynamical systems is stronger or not.
	In the next subsection, it will proven that for the control/reach problems,
	the bigger of the controllability region, the bigger the solution space of the input variables is, and then the better the some closed-loop control performance is. Thus, the size and shape of the controllability region $R_{a}$ is used to define and describe the control ability with time attribute.
	
	Next, 2 equivalent definitions on the stronger control ability between the two differential controlled plants, or of one controlled plants with two sets of systems parameters are purposed as follows.
	
	\begin{defi} \label {de:d0501}
		For the given controllability regions $R_{a}^{(1)}(T)$ and $ R_{a}^{(2)}(T)$ of the two LCT Systems $\Sigma_1$ and $\Sigma_2$,
		if the given $x_0 \in \partial R_{a}^{(1)}(T_1) \cap \partial R_{a}^{(2)}(T_2)$ and $T_1 < T_2$, the control ability of the systems $\Sigma_1$ at state $x_0$ is stronger than the systems $\Sigma_2$. If for any $T>0$, the control ability of the systems $\Sigma_1$ at all state in $\partial R_{a}^{(1)}(T) \cap \partial R_{a}^{(2)}(T)$ 
		is stronger than the systems $\Sigma_1$, the control ability of the systems $\Sigma_1$ is stronger than the systems $\Sigma_2$.
	\end{defi}
	
	\begin{defi} \label {de:d0502}
		For any $T>0$, if the two controllability regions $R_{a}^{(1)}(T)$ and $ R_{a}^{(2)}(T)$ of the LCT Systems $\Sigma_1$ and $\Sigma_2$ satisfy 
		\begin{align} 
		R_{a}^{(1)} (T) \supset R_{a}^{(2)}(T) \; \textnormal {and}\; \partial R_{a}^{(1)} (T) \cap \partial R_{a}^{(2)}(T) = \phi \label{eq:a05014},
		\end{align}
		the control ability of the systems $\Sigma_1$ is stronger than the systems $\Sigma_2$. if $R_{a}^{(1)}(T)$ and $ R_{a}^{(2)}(T)$ for any $T$ satisfy 
		\begin{align} 
		R_{a}^{(1)} (T) \supseteq R_{a}^{(2)}(T) \; \textnormal {and}\; \partial R_{a}^{(1)} (T) \cap \partial R_{a}^{(2)}(T) \neq \phi \label{eq:a05015},
		\end{align}
		the control ability of the systems $\Sigma_1$ is not weaker than the systems $\Sigma_2$.
	\end{defi}
	
	According to the above definitions, the control ability of the dynamical systems can be compared for many control engineering problems as stated above.
	
	\section{Theorem on Relation between the Open-loop Control Ability and the Closed-loop Performances}
	
Before discussing on relation between the open-loop control ability and the closed-loop performances, a time-optimal property for the boundary of controllability region is stated as follows \cite{Hulin2001}, \cite{Hu2002}.

	\begin{property}
		\label{pr:pr053}
		If the LCT Systems \eqref{eq:a0501} is controllable and the given state $x_0$ satisfy
		\begin{align} 
		x_{0} \in \partial R_{a} (T) \label{eq:a05016}
		\end{align}
		the time waste of the time-optimal control problem for stabilizing the initial state $x_0$ to the original of the state space (or for controlling the state at the original to the expecting state $x_0$ ) under the input amplitude constraint $U_a$ is $T$, \textit{that is}, the fewest control time is $T$.
	\end{property}

	Based on the definition of the control ability and above properties, a theorem on the
	relations among the open-loop control ability, the solution space of the input variables, and the closed-loop performances are purposed and proven as follows.
	
	\begin{theorem} \label{th:t0501} 
		It is assumed that two LCT Systems $\Sigma_1$ and $\Sigma_2$ are controllable, and their controllability regions are $R^{(1)}_{a} (T) $  and $R^{(2)}_{a} (T)$ respectively.
		If we have
		\begin{align} \label{eq:a05017}
		R^{(1)}_{a} (\tau) \subseteq R^{(2)}_{a} (\tau),\; \forall \tau \le T,
		\end{align}
		for the control problem stabilizing the state $x_0 \left (x_0 \in R^{(1)}_{a} (T) \cap R ^{(2)}_{a} (T) \right) $ to the original of the state space (or for controlling the state at the original to the expecting state $x_0$ ),
		the following conclusions hold under the input amplitude constraint $U_a$.
		
		1) The time waste of the time-optimal control for the system $\Sigma_2$ is not more than that of $\Sigma_1$, \textit{that is}, there exist some control strategies with the less control time and the faster response speed for the system $\Sigma_2$. 
		
		2) There exist more control strategies for the system $\Sigma_2$, \textit{that is}, 
		the bigger the controllability region is, the bigger the solution space of the input variable for the control problems, the easier designing and implementing the control are. 
	\end{theorem}
	
	The above theorem will be proven by the discretization model with a sufficient small sampling period for approximating the LCT systems. The discretization model with a small sampling period $\Delta$ for the LCT systems \eqref{eq:a0501} should be as follows \cite{Kailath1980} \cite{Chen1998}
			\begin{align} \label{eq:a05018}
	x_{k+1}=\hat A x_k+ \hat B u_k
	\end{align}
where $x_k$ and $u_k$ are the state variable and input variable of the discretization model, respectively.	
			\begin{align} \label{eq:a05019}
\hat A =e^{A\Delta}; \;\; \hat B =B \Delta
\end{align}
Therefore, when $\Delta \rightarrow 0$, the linear discrete-time (LDT) systems $\left( \hat A , \hat B \right)$ will approxiamte the LCT systems $(A,B)$.

	{\bfseries Proof of Theorem \ref{th:t0501} } Beacuse that the LCT systems $(A,B)$ can be approximated by the LDT systems $\left( \hat A , \hat B \right)$ with a sufficient small sampling-period $\Delta$, the conclussions of {\bfseries Theorem 1} in paper \cite {zhaomw202003} for the LDT systems are still established when $\Delta \rightarrow 0$ and then the conclussions in {\bfseries Theorem \ref{th:t0501} } for the LCT systems, similar to that {\bfseries Theorem 1} in paper \cite {zhaomw202003}, will be regarded as true.	
\qed

	By {\bfseries Theorem \ref{th:t0501}}, we have the following discussions:
	
	(1) Not only the time waste can be reduced by promoting the control ability, but also other closed-loop performance related the control time waste can be improved.
	
	(2) In fact, that the solution space of the input variable for the control problems is bigger implies that the control strategies in the solution space are with better robustness, and then the closed-loop control systems is also with better robustness. 
	
	(3) According to the theorem, the control ability defined and discussed here is indeed a control ability with time attributy. In fact, optimizing the control ability of the open-loop systems is to equal to promote some closed-loop performance indices related to the control time. 
	
	Therefore, optimizing the open-loop control ability with time attribute are with very greater signification and it's very necessary to optimize the control ability for these practical control engineering problems. To optimize the control ability, it is necessary to establish the quantify analysis and computing method for that. An analytical computing equation for the volume of the controllability region for the LDT systems is deduced in paper \cite{zhaomw202001} and the some analytical factors about the shape of the controllability region can be got by deconstructing the volume equation. 
	Next, the similar studies for the LCT systems are carried out and then based on these analytical expressions of the volume and shape factors, the optimizing and promoting methods for the control ability can be set up conveniently.
	
	\section {The Volume Computing for the Smooth Zonotope for the Matrix $A$ with $n$ Different Eigenvalues}

The volume computing problem for the zonotope generated by a matrix pair $(A,B)$ is discussed and two recurssive computing methods with the low time complexity are deduced in paper \cite {zhaomw202001}. Furthermore, an analytic volume-computing theorem about the zonotope that the matrix $A$ is with $n$ different real eigenvalues are purposed and proven. In paper \cite {zhaomw202004}, the analytic volume-computing theorem for that the matrix $A$ is with real repeated eigenvalues is proven also. Because that 
the narrow controllability region and reachability region can be described by the zonotope generated by the matrix pair, these results on the recurrsive and analytic volume computing also are used to analyze these regions and control abilities for the LDT systems. Later, these computing and analyzing methods for the LDT systems in papers \cite {zhaomw202001} \cite {zhaomw202004} wlii be generalized to the LCT systems.
	
	\subsection {the recurrsive computing of the volume of the smooth zonotope}

As the proof of \textbf{Theorem \ref{th:t0501}}, the LCT systems can be approximated by a LDT systems with a sufficient small sampling-period $\Delta$, and then the volume computing problem for the controllability region of the LCT systems can be transformated as that for LDT systems in papers \cite {zhaomw202001} \cite {zhaomw202004}.

It is assumed the sufficient small sampling-period is $\Delta$, the controllability region $R_{a}(T)$ defined in Eq. \eqref{eq:a0501002}  can be represented approximately as follows 
\begin{align} 
R_{a}(N) \approxeq \left\{x : x=\frac {\Delta}{2} \left[ Bu_{ 0} +2 \sum _{i=1}^{N-1} e^{A\Delta i} Bu_{ i} +e^{A\Delta N} Bu_{ N} \right], \;\; \forall \Vert u_i \Vert _\infty \le 1 \right\} \label{eq:a0501021} 
\end{align}
or
\begin{align} 
R_{a} (N) \approxeq \left\{x : x=\Delta \sum _{i=0}^{N-1} e^{A\Delta i} Bu_{ i} , \;\; \forall \Vert u_i \Vert _\infty \le 1 \right\} \label{eq:a0501022} 
\end{align}
where $N$ is the integer part of $T/\Delta$. Therefore, according to the above approximation equations, the controllability region $R_{a}(T)$ can be regarded as a zonotope $R_a(N)$ generated by the matrix pair $\left( \hat A, \hat B \right)$ as Eq. \eqref{eq:a05019}, and then the recurssive volume-computing methods for the controllability regions fo the LDT systems in \textbf {Section 2} and \textbf {Section 3} of paper \cite {zhaomw202001} can be applied to that of the LDT systems.

	\subsection {the analytical computing of the volume of the time-infinite smooth zonotope}
			In paper \cite{zhaomw202001}, an analytic volume-computing equation for the zonotope $R_a (N)$ generated by the matrix pair $(A,B)$ with $n$ different real eigenvalues of matrix $A$ is proven. In fact, the region $R_a (N)$ defined by Eq. \eqref{eq:a0501022}
	is a special zonotope defined in paper \cite{zhaomw202001}. Therefore, the quantifying and maximizing of the controllability region $R_a (N)$ can be carried out based on the results in paper \cite{zhaomw202001}. The results in the volume computation of the zonotope generated by the matrix pair $(A,B)$, in \textbf{Theorem 2} and \textbf{Theorem 3} of paper \cite{zhaomw202001} can be summarized and generalized to the LCT systems. Hence, we have the following theorem about the analytical volume computing for the infinite-time controllability region $R_a(\infty)$ of the LCT systems.
	
	\begin{theorem} \label {th:t0502} If matrix $A\in R^{n\times n}$ is with $n$ different eigenvalues in the interval $(-\infty,0)$ and $b \in R^n$, the volume of the infinite-time smooth zonotope $R_a (\infty)$ generated by matrix pair $ ( A, B) $ can be computed analytically by the following equation:
		\begin{equation} \label{eq:as17}
		\textnormal {vol} (R_\infty) =2^n \left| \det(P) \left( \prod_{1 \leq j_{1}<j_{2} \leq n} \frac{ \lambda_{j_{2}}- \lambda_{j_{1}}}{ \lambda_{j_{1}}+ \lambda_{j_{2}}} \right) \left( \prod_{i=1}^{n} \frac{q_{i} b}{ \lambda_{i}} \right) \right| 
		\end{equation}
		where $\lambda_{i}$ and $q_{i}$ are the $i$-th eigenvalue and the corresponding unit left eigenvector of matrix $A$, matrix $P$ is the matrix transforming the matrix $A$ as a diagonal matrix.
	\end{theorem} 

\textbf{Proof of Theorem \ref{th:t0502}} Similar to above approximation method by the sampling model, the smooth zonotope generated by the matrix pair $(A,b)$ can be approximated by the zonotope generated by the matrix pair $\left( \hat A, \hat b \right)$ as Eq. \eqref{eq:a05019}. When the matrix $A$ is with $n$ different eigenvalues $\lambda _i $ in interval $(-\infty,0)$, the matrix $\hat A=e^{A\Delta}$ is also with $n$ different eigenvalues $e^{ \Delta \lambda_i}$ in interval $[0,1)$. And then, by the \textbf{Theorem 2} in paper \cite{zhaomw202001}, we have the following volume computation of the zonotope generated by $\left( \hat A, \hat b \right)$ as follows.
\begin{equation} \label{eq:as171}
\textnormal {vol} (R_\infty) \approxeq 2^n \left| \det(P) \left( \prod_{1 \leq j_{1}<j_{2} \leq n} \frac{ e^{ \Delta \lambda_{j_2}} - e^{ \Delta \lambda_{j_1}} } {1- e^{ \Delta \left( \lambda_{j_1} + \lambda_{j_2} \right) } } \right) \left( \prod_{i=1}^{n} \frac{q_{i} b \Delta}{1- e^{ \Delta \lambda_{i}} } \right) \right| 
\end{equation}
So, when $\Delta \rightarrow 0$, we have
		\begin{equation} \label{eq:as172}
\textnormal {vol} (R_\infty) = 2^n \left| \det(P) \left( \prod_{1 \leq j_{1}<j_{2} \leq n} \frac{ \lambda_{j_{2}}- \lambda_{j_{1}}}{ \lambda_{j_{1}}+ \lambda_{j_{2}}} \right) \left( \prod_{i=1}^{n} \frac{q_{i} b}{ \lambda_{i}} \right) \right| 
\end{equation}
	\textit{that is}, the theorem is true.
	\qed
	
	Based on the above theorem, the volume of the controllability region $R_a (N)$ when $N \rightarrow \infty$ can be computed analytically.
	Furthermore, some analytical factors describing the shape of the $R_a (N)$ can be got by deconstructing the analytical volume computing equation \eqref{eq:as17}. These analytical expressions for these volume and shape factors can be describe quantitatively the control ability of the dynamical systems, and then optimizing these volume and shape factors is indeed maximizing the control ability. 
	
	\subsection{Decoding the Controllability Region}
	
	According to the volume computing equation \eqref{eq:as17}, some factors described the shape and size of the controllability region, \textit{that is}, the control ability of the dynamical systems, are deconstructed as follows.
	\begin{align} 
	F_1 & = \left \vert \prod_{1 \leq j_{1}<j_{2} \leq n} \frac{ \lambda_{j_{2}}- \lambda_{j_{1}} }{ \lambda_{j_{1}} + \lambda_{j_{2}}} \right \vert \label{eq:630} \\
	F_{2,i} 
	&= \frac{ \left \vert q_ib \right \vert}{ \lambda_{i} } 
	\label{eq:640} \\
	F_{3,i} & =\left \vert q_ib \right \vert \label{eq:650}
	\end{align}
	The above analytical factors can be called respectively as 
	the shape factor, the side length of the circumscribed rhombohedral, and the modal controllability. In fact, the shape factor $F_1$ is also the eigenvalue evenness factor of the linear system, and can describe the control ability caused by the eigenvalue distribution. In addition, the modal controllability factor $F_{3,i}$ have been put forth by papers \cite{chan1984} \cite{HamElad1988} \cite{ChoParLee2000} \cite{Chenliu2001}, and will not be discussed here.
	
	\subsubsection {The Shape Factor of the Reachability Region and the Eigenvalue Evenness Factor of the Linear System}
	
\begin{figure}[htbp]
		\centering
		\begin{minipage}[c]{0.45\textwidth} %minipage使之保持同一行,0.2占这行的0.2
			\centering
			\includegraphics[width=0.8\textwidth]{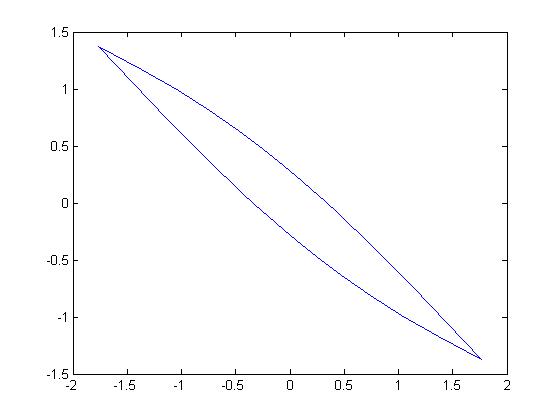} %0.5指图片宽度
			%			\caption[c]{ $F_1=0.7813$ $\lambda_1,\lambda_2$=0.4,0.9 \notag \label {fig:as01}}
			\\ (a) \footnotesize {(-2.9,-0.75,0.5891)}
		\end{minipage}%	
		\begin{minipage}[c]{0.45\textwidth}
			\centering
			\includegraphics[width=0.8\textwidth]{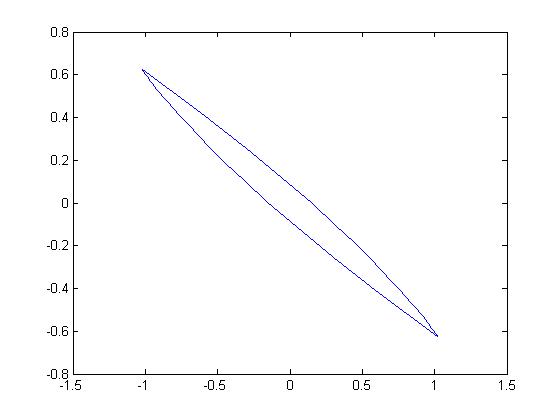}
			%			\caption[c]{$F_1=0.6522$ \label {fig:as02}}
			\\ (b) \footnotesize {(-2.9,-1.75,0.2473)}
		\end{minipage}
		\begin{minipage}[c]{0.45\textwidth}
			\centering
			\includegraphics[width=0.8\textwidth]{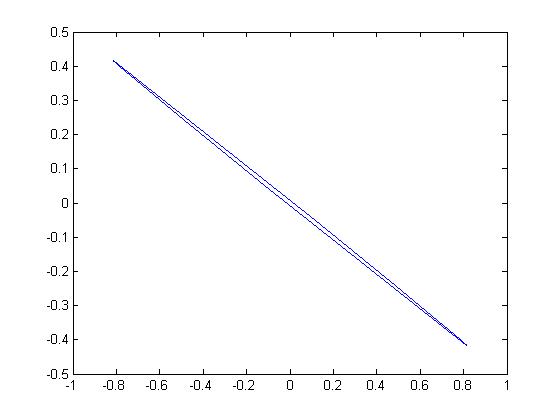}
			%			\caption{$F_1=0.2128$ \label {fig:as03}}
			\\ (c) \footnotesize {(-2.9,-2.75,0.0265)}
		\end{minipage}
		\caption[c]{The 2-dimensional smooth zonotope with $(\lambda_1,\lambda_2,F_1)$ \label {fig:as01}}	
	\end{figure}

\begin{figure}[htbp]
		\centering
		\begin{minipage}[c]{0.45\textwidth} %minipage使之保持同一行,0.2占这行的0.2
			\centering
			\includegraphics[width=0.8\textwidth]{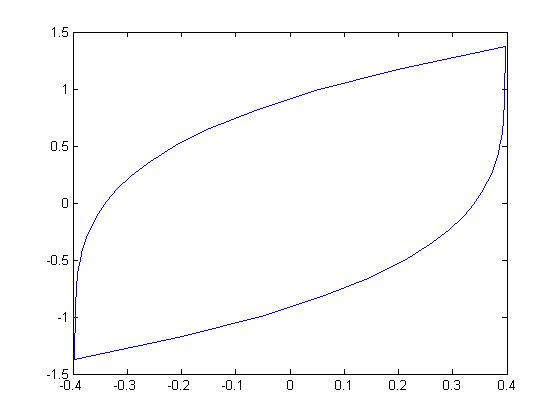} %0.5指图片宽度
			%			\caption[c]{ $F_1=0.7813$ $\lambda_1,\lambda_2$=0.4,0.9 \notag \label {fig:as01}}
			\\ (a) \footnotesize {(-2.9,-0.75,0.5891)}
		\end{minipage}%	
		\begin{minipage}[c]{0.45\textwidth}
			\centering
			\includegraphics[width=0.8\textwidth]{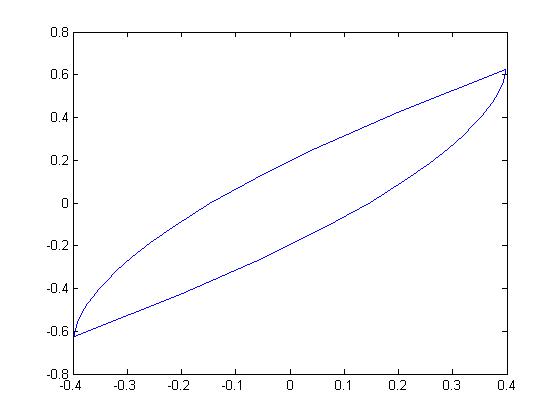}
			%			\caption[c]{$F_1=0.6522$ \label {fig:as02}}
			\\ (b) \footnotesize {(-2.9,-1.75,0.2473)}
		\end{minipage}
		\begin{minipage}[c]{0.45\textwidth}
			\centering
			\includegraphics[width=0.8\textwidth]{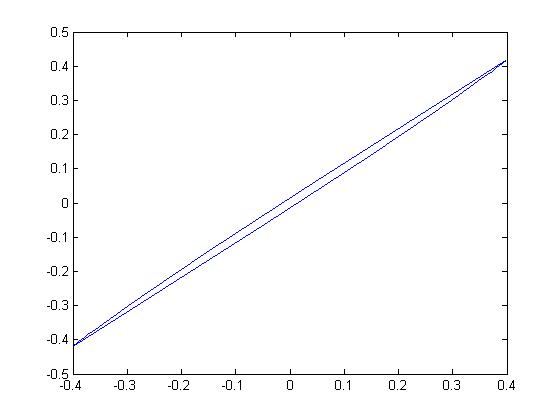}
			%			\caption{$F_1=0.2128$ \label {fig:as03}}
			\\ (c) \footnotesize {(-2.9,-2.75,0.0265)}
		\end{minipage}
		\caption[c]{The 2-dimensional smooth zonotope with $(\lambda_1,\lambda_2,F_1)$ \label {fig:as02}}	
	\end{figure}

	Fig. \ref{fig:as01} shows the 2-dimensional smooth zonotopes $R_a (6s) $, \textit{i.e.}, the terminal time $ T=6s$, generated by the 3 matrix pairs $(A,b)$ that the matrix $A$ is with the different eigenvalues and matrix $b$ is a same vector, and Fig. \ref{fig:as02} shows the 2-dimensional smooth zonotopes generated by the diagonal matrix pairs of these 3 matrix pairs $(A,b)$, \textit{that is}, the smooth zonotopes in Fig. \ref{fig:as02} are in the invariant eigen-space.

	%\caption[short text]{text}
	
	From these figures, we can see, when the two eigenvalues of the matrix $A$ are approximately equal, the minimum distances of the boundary of the region $R_a (T)$ to the original will be approximately zero, and the region $R_a (T)$ will be flattened. Similar casees are also for the $n$-dimensional smooth zonotope generated by the matrix pair.
	Therefore, the distributions of all eigenvalues of the matrix $A$ are even, the ratio between the minimum and maximum distance of the boundary of the smooth zonotope generated by the pair $(A,b)$ can be avoided as a small value and the smooth zonotope will be avoided flattened. 
	
	The factor $F_1$ deconstructed from the volume computing equation \eqref{eq:as17} can be used to describe the uniformity of the $n$ biggest distances of the region $ R_a(T)$ in $n$ eigenvectors. The bigger the value of the factor $F_1$, the bigger 
	the ratio between the minimum and maximum distances of the bounded of the region $\overline R_a(T)$ is, and then the greater the volume of the region is.
	
	Otherwise, the factor $F_1$ can be used to describe the evenness of the eigenvalue distribution of the linear system $\Sigma(A,B)$. The bigger the value of the factor $F_1$, the bigger the controllable region of the system is, and the stronger the control ability of the systems is.
	
	\subsubsection {The circumscribed hypercube and circumscribed rhombohedral of the controllability region}
	
	The factor $F_{2,i}$ is indeed the biggest distance of the region $ R_N^d$ in the $i$-dimensional coordinate of the eigen-space (shown as in Fig. \ref{fig:as02} ), \textit{that is}, the $n$ side lengths of the circumscribed hypercube of the region $R_a {T}$ are $2F_{2,i},i=\overline{1,n}$. By the volume equation \eqref{eq:as17}, the volume of the region can be represented as the production of the volume $\prod _{i=1} ^{n} F_{2,i}$ of the circumscribed hypercube and the shape factor $F_1$
	
	Because that these expressions of the volume and the shape factors can describe accurately the size and shape of the smooth zonotope generated by the matrix pair, \textit{i.e.}, the control ability of the dynamical systems \cite{zhaomw202003}, these expressions can be used conveniently to be the objective function or the constrained conditions for the optimizing problems of the control ability of dynamical systems.

	\section{The Control Ability for the Systems with the Repeated real Eigenvalues}
	
	\subsection{A Property about the Control Ability for the Systems with the Repeated Eigenvalues}
	
	In the last section, the volume of the smooth zonotope generated by the matrix pair $(A,b)$, correspondingly the control ability of the linear systems, is discussed for the matrix $A$ with $n$ different real eigenvalues. Next, the volume is discussed for that the matrix $A$ is with the repeated real eigenvalues. 
	
	When some eigenvalues of matrix $A$ are the repeated eigenvalues, the matrix $A$ can be transformed as a Jordan matrix by a similar transformation in the matrix theory. In view of this, the following discussion are only carried out for the Jordan matrix. For the volume of the smooth zonotope generated by the Jordan matrix pair $(A,b)$, a following theorem about the relation between the volume and matrix $b$ can be stated and proven firstly.
	
	\begin{theorem} \label{th:t0503}
		The volume of the smooth zonotope generated by the Jordan matrix pair $(A,b)$ has relation only to the last rows of the matrix blocks of the matrix $b$, corresponding to the each Jordan block of the matrix $A$, and has no relation to other rows.
	\end{theorem}
	
	The conclusion in \textbf{Theorem \ref{th:t0503}} is consistent with the conlusion, in control theory, that 
	the state controllability of the systems $ \Sigma(A,b)$ has relation only to the last rows of these matrix blocks of the matrix $b$, corresponding to the each Jordan block of the Jordan matrix $A$, and has no relation to other rows. According to \textbf{Theorem \ref{th:t0503}}, for simplifying the volume computation for the Jordan matrix pair $(A,b)$, the rows of matrix $b$ entirely unrelated to the volume will be regarded as 0.
	
	\textbf{Proof:} Without loss of the generality, the theorem is proven only for the Jordan matrix with one Jordan block, and other cases can be proven similarly.
	
	When the matrix $A$ is with only one Jordan block, the two matrices of the matrix pair $(A,b)$ can be written as follows
	\begin{equation} \label{eq:c3b01}
	A= \left[ \begin{array}{ccccc}
	\lambda & 1 & 0 & \cdots & 0\\
	0 & \lambda & 1 & \cdots & 0\\
	0 & 0 & \lambda & \cdots & 0\\
	\vdots & \vdots & \vdots & \ddots & \vdots\\
	0 & 0 & 0 & \cdots & \lambda
	\end{array} \right], \quad b= \left[ \begin{array}{c}
	b_{1}\\
	b_{2}\\
	b_{3}\\
	\vdots\\
	b_{n}
	\end{array} \right]
	\end{equation}
	the corrsponding two matrices of the matrix pair $ \left(\hat A, \hat b\right)$ of the sampling model as follows
	\begin{equation} \label{eq:c3b012}
			\hat A=e^{A\Delta}= e^{\lambda \Delta}\left[ \begin{array}{ccccc}
		1 & \Delta & \frac {\Delta^2}{2} & \cdots & \frac {\Delta^{m_i-1} }{ \left( m_i-1 \right) ! }\\
		0 & 1 & \Delta & \cdots & \frac {\Delta^{m_i-2} }{ \left( m_i-2 \right) ! }\\
		0 & 0 & 1 & \cdots & \frac {\Delta^{m_i-3} }{ \left( m_i-3 \right) ! }\\
		\vdots & \vdots & \vdots & \ddots & \vdots\\
		0 & 0 & 0 & \cdots & 1
	\end{array} \right], \quad 
	\hat b=b\Delta= \Delta \left[ \begin{array}{c}
		b_{1}\\
		b_{2}\\
		b_{3}\\
		\vdots\\
		b_{n}
	\end{array} \right]
\end{equation}
And then, the matrix transformated the matrix $\hat A$ as a Jordan matrix is as follows

	\begin{equation} \label{eq:c3b013}
P= \left[ \begin{array}{ccccc}
1 & 0 & 0 & \cdots & 0 \\
0 & \Delta^{-1} e^{-\lambda \Delta} & * & \cdots & * \\
0 & 0 & \Delta^{-2} e^{-2\lambda \Delta} & \cdots & * \\
\vdots & \vdots & \vdots & \ddots & \vdots\\
0 & 0 & 0 & \cdots & \left( \Delta e^{- \lambda \Delta} \right) ^{-m_i+1}
\end{array} \right]
\end{equation}
where '*' means some value. Therefore, we have
	\begin{equation} \label{eq:c3b013z}
P^{-1}= \left[ \begin{array}{ccccc}
1 & 0 & 0 & \cdots & 0 \\
0 & \Delta e^{\lambda \Delta} & * & \cdots & * \\
0 & 0 & \Delta^{2} e^{2\lambda \Delta} & \cdots & * \\
\vdots & \vdots & \vdots & \ddots & \vdots\\
0 & 0 & 0 & \cdots & \left( \Delta e^{\lambda \Delta} \right) ^{m_i-1}
\end{array} \right]
\end{equation}
and the corresponding matrices of the Jordan matrix pair are as follows
	\begin{align} \label{eq:c3b01z}
\overline A &= P^{-1} \hat AP=\left[ \begin{array}{ccccc}
e^{\lambda \Delta} & 1 & 0 & \cdots & 0\\
0 & e^{\lambda \Delta} & 1 & \cdots & 0\\
0 & 0 & e^{\lambda \Delta} & \cdots & 0\\
\vdots & \vdots & \vdots & \ddots & \vdots\\
0 & 0 & 0 & \cdots & e^{\lambda \Delta}
\end{array} \right], \quad \\
\overline b & =P^{-1} \hat b= \left[ \begin{array}{c}
*\\
*\\
*\\
\vdots\\
 \left( \Delta e^{\lambda \Delta} \right) ^{m_i-1} b_{n}
\end{array} \right]
\end{align}

Therefore, by \textbf{Theorem 3} in paper \cite {zhaomw202004}, the volume of the zonotope 
	 generated by the Jordan matrix pair $\left ( \overline A, \overline b \right)$ has relation only to the last rows of the matrix blocks of the matrix $\overline b$, corresponding to the each Jordan block of the matrix $\overline A$, and has no relation to other rows, \textit{that is}, by Eq. \eqref{eq:c3b012}, we can see, the approximating-zonotope volume $R_a(N)$ is only related to parameter $b_n$ of the LCT systems $(A,b)$ and not to other $b_i$. Thus, considered that the LDT systems $\left ( \overline A, \overline b \right)$ will be equal to the LCT systems $(A,b)$ when $\Delta \rightarrow 0$, we know, 
	 the conclusion of \textbf{Theorem \ref{th:t0503}} holds.
\qed
	
	According to the theorem, the volume computation of the smooth zonotope for the Jordan matrix pair $(A,b)$ can be equivalent to the volume computation for the Jordan matrix pair $(A, \beta)$, where $ \beta=[0, \dots,0, b_n]^T$.

	\subsection{Volume Computing for the Infinite-time Smooth Zonotope When the Jordan Matrix $A$ with Multiple Jordan Blocks}
	
	If the matrix $A$ is a Jordan matrix with multiple Jordan blocks, the Jordan matrix pair $(A,b)$ can be denoted by
	\begin{align}
	(A,b)=\left( \textnormal{diag-block} \left\{ J_1, J_2, \dots, J_q \right\} ,
	\left [\beta_1^T, \beta_2^T, \dots, \beta_q^T \right]^T \right) \label {eq:as002}
	\end{align}
	where $q$ is the Jordan block number, $J_i$ is a $m_i \times m_i$ Jordan block with the eigenvalue $\lambda_i$, the matrix block $\beta_i$ is $\left [b_{i,1}, b_{i,2}, \dots, b_{i,m_i} \right]^T$. And then, we have
	$$ \sum _{i=1} ^{q} m_i =n $$
	
	Similar to \textbf{Theorem 5} for the LDT Jordan models with multiple Jordan blocks  in paper \cite{zhaomw202004}, a theorem about the volume computation of the infinite-time smooth zonotope for the LCT Jordan models with multiple Jordan blocks can be stated as follows.
	
	\begin{theorem} \label{th:t0505} When all eigenvalues of the matrix $A$ satisfy $ \lambda \in (- \infty,0) $, the volume of the infinite-time smooth zonotope generated by the Jordan matrix pair $(A,b)$ as Eq. \eqref {eq:as002} is computed as follows 
		\begin{align} 
		V = 2^n \left \vert \prod _{i=1}^{q-1} \prod _{j=i+1}^{q} \left( \frac{\lambda_i - \lambda_j}{ \lambda_i + \lambda_j } \right)^{m_i \times m_j} \right \vert
		\times	
		\left \vert \prod _{i=1}^{q} \frac{	\left(b_{i,m_i}/ \lambda_i \right) ^{m_i} } { \left( 2\lambda_i \right) ^{m_i(m_i-1)/2}}
		\right \vert \label{eq:e3d0192} 
\end{align}
		When matrix $A$ isn't a Jordan matrix, the volume of the infinite-time smooth zonotope generated by the matrix pair $(A,b)$ is computed as follows 
		\begin{equation} \label{eq:e3d01921}
		V = 2^n \left \vert \prod _{i=1}^{q-1} \prod _{j=i+1}^{q} \left( \frac{\lambda_i - \lambda_j}{ \lambda_i +\lambda_j } \right)^{m_i \times m_j} \right \vert
		\times	
		\left \vert \det \left (P_J \right) \prod _{i=1}^{q} \frac{	\left(b_{i,m_i}/ \lambda_i \right) ^{m_i} } { \left( 2\lambda_i \right) ^{m_i(m_i-1)/2}}
		\right \vert
		\end{equation}
		where the matrix $P_J$ is the Jordan transformation matrix and the vector $q_i$ is the only unit left eigenvector of the matrix $A$ for the eigenvalue $\lambda_i$.
	\end{theorem}	
	
	Based on \textbf{Theorem 5} in paper \cite{zhaomw202004}, the above theorem can be proven as the proof of \textbf{Theorem \ref{th:t0502}} by the discreting approximating models.

	\subsection{Decoding the Volume of the Controllability Region}
	
	According to the computing equation \eqref{eq:e3d01921}, some factors described the shape and size of the controllability religion, \textit{that is}, the smooth zonotope generated by the matrix pair $(A,B)$ with the repeated eigenvalues, are deconstructed as follows.
	\begin{align} 
	F_1 & = \left \vert \prod _{i=1}^{q-1} \prod _{j=i+1}^{q} \left( \frac{\lambda_i - \lambda_j}{ \lambda_i + \lambda_j } \right)^{m_i \times m_j} \right \vert
	\times	\left \vert \prod _{i=1}^{q} \frac{	1 } { \lambda_i ^{m_i(m_i-1)/2}} \right \vert
	\label{eq:63} \\
	F_{2,i,j} 
	&= \left \{ 
	\begin{array}{ll}
	\frac{ \left \vert q_{i,j}b \right \vert}{ \lambda_{i} }
	& j=m_i \\
	\frac{ \left \vert q_{i,j}b +F_{i,j+1} \right \vert}{ \lambda_{i} } & j=m_i-1,m_i-2,\dots
	\end{array}
	\right. \;\; i=1,2,\dots,q
	\label{eq:64} \\
	F_{3,i} & =\left \vert q_{i,m_i}b\right \vert^{m_i}
	\end{align}
	where all $ q_{i,j}b$ are assumed as with the same sign. These 3 classes of factors are similar to the last section, called as the shape factor, the side length of the circumscribed rhombohedral, and the modal controllability. These factors can be describe the shape and size of the smooth zonotope, the control ability of the system, and the eigenvalue evenness factor of the linear system.
	
	\subsubsection {The Smooth Zonotope Shape Factor and the Eigenvalue Evenness Factor of the Linear System}
	
	Similar to the analysis for the matrix $A$ with $n$ different eigenvalues in last section, by Eq. \eqref{eq:e3d01921}, we can see, 
	when some two eigenvalues of the two Jordan blocks of the system matrix $A$ are approximately equal, the minimum distance of the boundary of the smooth zonotope to the original of the stat space will be approximately zero, and the smooth zonotope will be flattened.
	Therefore, the distributions of all eigenvalues of the matrix $A$ are even, the ratio between the minimum and maximum distance of the boundary to original can be avoided as a small value and the smooth zonotope, \textit{that is}, the control region, will be avoided flattened. And then, the volume of the smooth zonotope and the control ability will maintain a certain size.
	
	The factor $F_1$ deconstructed from the volume computing equation \eqref{eq:e3d01921} can be used to describe the uniformity the $n$ distance of the boundary to the original in the $n$ eigenvector. The bigger the value of the factor $F_1$, the bigger 
	the ratio between the minimum and maximum distance of the boundary is, and then the greater the volume of the smooth zonotope.
	
	Otherwise, the factor $F_1$ can be used to describe the evenness of the eigenvalue distribution of the linear system $\Sigma(A,B)$. The bigger the value of the factor $F_1$, the bigger the controllable region of the system is, and the stronger the control ability of the systems is.
	
	\subsubsection {The circumscribed hypercube and circumscribed rhombohedral of the controllability region}
	
\begin{figure}[htbp]
		\centering
		\includegraphics[width=0.8\textwidth]{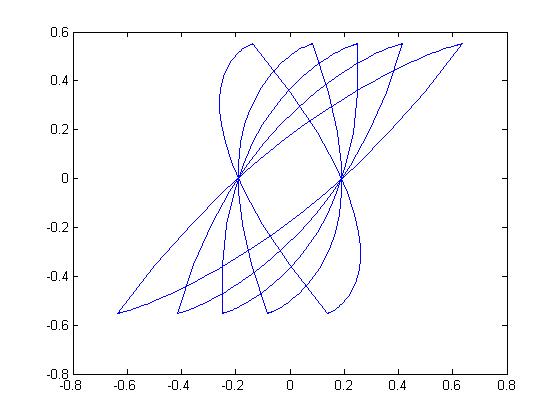} %0.5指图片宽度
		\caption[c]{The 3 2-D smooth zonotopes for Jordan matrix pairs $(A,b)$ with the different $b$ \label {fig:as03}}	
	\end{figure}

	Fig. \ref{fig:as03} shows the 2-dimensional smooth zonotopes $R_a(6s)$ by the 3 Jordan matrix pairs $(A,b)$ as follows
	$$
	A=\left[ \begin{array}{cc}
	-2 & 1 \\
	0 & -2
	\end{array} \right] \;\; 
	b=\left[ \begin{array}{c}
	0.7 \\ 1
	\end{array} \right],\;
	\left[ \begin{array}{c}
	0.3 \\ 1
	\end{array} \right],\;
	\left[ \begin{array}{c}
	0 \\ 1
	\end{array} \right] , \; 
\left[ \begin{array}{c}
-0.3 \\ 1
\end{array} \right],\;
	\left[ \begin{array}{c}
	-0.7 \\ 1
	\end{array} \right] 
	$$
	
	By Fig. \ref{fig:as03}, we know, the factor $F_{2,i,j}$ is indeed the biggest distance of the boundary of the smooth zonotope in the each eigenvector, \textit{that is}, the $n$ side lengths of the circumscribed hypercube of the smooth zonotope in the invariant eigen-space are $2F_{2,i,j},i=\overline{1,n}$. By the volume equation \eqref{eq:e3d01921}, the volume of the smooth zonotope region can be represented as the production of the volume $\prod _{i=1} ^{n} F_{2,i,j}$ of the circumscribed hypercube and the shape factor $F_1$
	
	By \textbf {Theorem, \ref{th:t0503}}, we know, the volume of the smooth zonotope generated by the Jordan matrix pair $(A,b)$ has relation only to the last rows of the matrix blocks of the matrix $b$, corresponding to the each Jordan block of the matrix $A$, and has no relation to other rows.
	But from Fig. \ref{fig:as03}, we know, these 'other' rows of matrix $b$ of Jordan matrix pair $(A,b)$ don't affect the size of the volume but maybe affect the shape of the corresponding smooth zonotopes.
	
	\section {Numerical Experiments (It's not available for now)}
	
		\section {Conclusions}

			 In this paper, the control ability with time attributy for the linear conti- nuous-time (LCT) systems are defined and analyzed by the volume computing for the controllability region. Firstly, a relation theorem about the open-loop control ability, the control strategy space (\textit{i.e.}, the solution space of the input variable for control problems), and the some closed-loop performance for the LCT systems is purposed and proven. This theorem shows us the necessity  to optimize the control ability for the practical engineering problems. Secondly, recurssive  volume-computing algorithms with the low computing complexities for the finite-time controllability region are discussed. Finally, two analytical volume computations of the infinite-time controllability region for  the systems with $n$ different and repeated real eigenvalues are deduced, and then by deconstructing the volume computing equations, 3 classes of the shape factors are constructed. These analytical volume and shape factors can describe accurately the size and shape of the controllability region.
Because the time-attribute control ability for LCT systems is directly related to the controllability region with the unit input variables (\textit{i.e.} the input variable is with bounded value as 1), based on these analytical expressions on the volume and shape factors, the time-attribute control ability can be computed and optimized conveniently. And then, a noval researching field about the optimizing open-loop control ability and promoting the closed-loop control performance and robustness is expected to be founded in control theory and engineering. 
	
	\bibliographystyle{model1b-num-names}

%	\bibliography{zzz}
\end{document}